\documentclass[twocolumn,apl,showpacs,showkeys,preprintnumbers,amsmath,amssymb]{revtex4}
\usepackage{graphicx}
\usepackage{dcolumn}
\usepackage{bm}

\begin{document}

\title{Enhanced thermopower in an intergrowth cobalt oxide Li$_{0.48}$Na$_{0.35}$CoO$_{2}$}

\author{Zhi Ren,$^{1}$ Jingqin Shen,$^{1}$ Shuai Jiang,$^{1}$ Xiaoyang Chen,$^{1}$ Chunmu Feng,$^{2}$ Zhu'an Xu,$^{1}$ Guanghan Cao$^{1}$\footnote[1]{To whom correspondence should be addressed
(ghcao@zju.edu.cn)}}

\affiliation{$^{1}$Department of Physics, Zhejiang University,
Hangzhou 310027, People's Republic of China}

\affiliation{$^{2}$Test and Analysis Center, Zhejiang University,
Hangzhou 310027, People's Republic of China}

\date{\today}

\begin{abstract}
We report the measurements of thermopower, electrical resistivity
and thermal conductivity in a complex cobalt oxide
Li$_{0.48}$Na$_{0.35}$CoO$_{2}$, whose crystal structure can be
viewed as an intergrowth of the O3 phase of Li$_{x}$CoO$_{2}$ and
the P2 phase of Na$_{y}$CoO$_{2}$ along the $c$ axis. The compound
shows large room-temperature thermopower of $\sim$180 $\mu$V/K,
which is substantially higher than those of Li$_{x}$CoO$_{2}$ and
Na$_{y}$CoO$_{2}$. The figure of merit for the polycrystalline
sample increases rapidly with increasing temperature, and it
achieves nearly 10$^{-4}$ K$^{-1}$ at 300 K, suggesting that
Li$_{x}$Na$_{y}$CoO$_{2}$ system is a promising candidate for
thermoelectric applications.
\end{abstract}



\maketitle

Thermoelectric material enables direct and reversible
heat-to-electricity conversion, which can be applied for the clean
power generation and refrigeration only using solid-state
elements.\cite{Mahan,Tritt} The efficient and desirable
thermoelectric materials need not only to have high dimensionless
figure of merit defined by $ZT$=$S^{2}T$/$\rho$$\kappa$, where $S$
is the Seebeck coefficient or the thermopower, $T$ the absolute
temperature, $\rho$ the electrical resistivity, and $\kappa$ the
thermal conductivity, but also to be stable in air at elevated
temperatures. Conducting oxides have the advantage for the latter
aspect, however, they have long been ignored as potential
thermoelectric materials until the discovery of large thermopower
in NaCo$_{2}$O$_{4}$ by Terasaki $et$ $al.$\cite{Terasaki} The
material exhibits rather low $\rho$ as well as large $S$ (one
order of magnitude larger than in typical metals) at room
temperature. The power factor $S^{2}$/$\rho$ at 300 K is
comparable to the well-known thermoelectric material
(Bi$_{x}$Sb$_{1-x}$)$_2$Te$_3$\cite{Caillat}. The $ZT$ parameter
at 800 K was reported to exceed the criterion value of 1.0,
suggesting that the cobaltite be promising material for the
application of high-temperature thermoelectric power
generation.\cite{Fujita}

After the discovery in the NaCo$_{2}$O$_{4}$
system\cite{Terasaki}, a number of lamellar cobalt oxides
($e$.$g.$, Ca$_3$Co$_4$O$_9$\cite{Masset,Li,Miyazaki},
Bi$_2$Sr$_2$Co$_2$O$_y$\cite{Funahashi,Itoh}, and
TlSr$_2$Co$_2$O$_y$\cite{Hebert}) containing CdI$_{2}$-type
CoO$_{2}$ layers were found to display considerably high
thermopower. Therefore, the triangular CoO$_{2}$ layers reasonably
play an important role for the thermoelectric property. In other
word, it is rational to look for new thermoelectric oxides in the
system containing CoO$_{2}$ layers.

Balsys and Davis\cite{Balsys94} first reported a new layered
cobalt oxide Li$_{0.43}$Na$_{0.36}$CoO$_{1.96}$ in 1994. The
crystal structure can be described as an intergrowth of the O3
phase of Li$_{x}$CoO$_{2}$ (LCO)\cite{Mizushima} and the P2 phase
of Na$_{y}$CoO$_{2}$ (NCO)\cite{Fouassier} along the $c$ axis, as
revealed by the high resolution neutron powder diffraction
study\cite{Balsys94}. In between the edge-sharing CoO$_6$
octahedron layers, as seen in Fig. 1, the lithium ions are sited
within an octahedral oxygen framework whereas the sodium ions are
within a trigonal prismatic oxygen environment. Accordingly, the
structure of Li$_{x}$Na$_{y}$CoO$_{2}$ (LNCO) can be called OP4,
where the letter O represents the octahedral coordination for
lithium ions, the letter P the prismatic coordination for sodium
ions, and the digit 4 denotes four CoO$_{2}$ layers per unit cell.
Since the conductivity of Li$_{x}$CoO$_{2}$ is much lower than
that of Na$_{y}$CoO$_{2}$ when $x \sim y$\cite{Levasseur}, LNCO
can be made and adjusted to be a natural quantum-well-like
superlattice, which may possibly enhance the thermoelectric
performance\cite{Harman}. In this Letter, we report the
thermoelectric properties in this novel compound.

\begin{figure}[tbp]
\includegraphics[width=7cm]{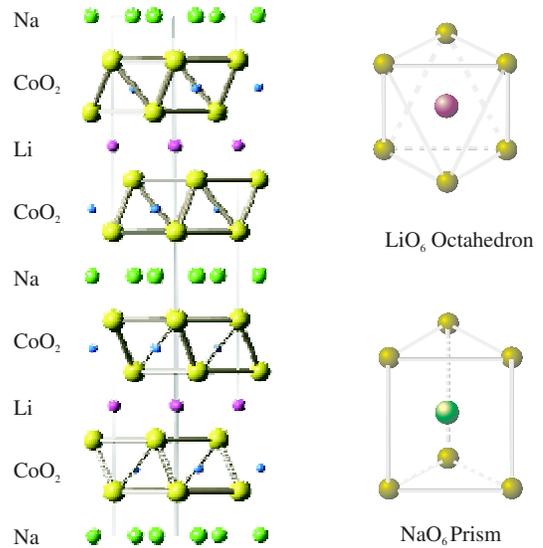}
\caption{(color online) Crystal structure of
Li$_{x}$Na$_{y}$CoO$_{2}$. The local oxygen environments of
lithium and sodium are shown on the right.}
\end{figure}

Balsys and Davis\cite{Balsys94} prepared the LNCO sample by
several steps. They prepared Na$_{0.7}$CoO$_{2}$ and LiCoO$_{2}$
in advance, then mixed the two compounds in equal molar ratio. The
mixture was finally fired at 1123 K for 4 days in an air
atmosphere. The report implies that the resultant sample contained
some LCO as a secondary phase. We also found that this kind of
preparation led to some impurities of LCO and/or NCO.
Alternatively, we synthesized the LNCO samples by direct
solid-state reaction using high-purity Li$_{2}$CO$_{3}$,
Na$_{2}$CO$_{3}$, Co$_{3}$O$_{4}$ as the starting materials. The
well-mixed powders with different stoichiometry were pressed into
pellets under a pressure of 2000 kg/cm$^{2}$. The pellets were
rapidly heated to 1173 K in flowing oxygen, holding for 30 hours,
and then quenched. The as-quenched samples were finally annealed
in an oxygen flow at 673 K for 24 hours and allowed to cool down
to room temperature slowly.

Powder x-ray diffraction (XRD) experiments were carried out using
a D/Max-$\gamma$A diffractometer with the Cu K$\alpha$ radiations.
It was found that single-phase LNCO samples could be formed only
in a narrow range of the composition. One of our best samples was
obtained with the initial composition of
Li$_{0.47}$Na$_{0.44}$CoO$_{2}$. By the inductively coupled plasma
atomic-emission spectrometry (ICP-AES) measurement, the final
composition was determined to be
Li$_{0.48(1)}$Na$_{0.35(1)}$CoO$_{2}$, assuming no oxygen
deficiencies. The difference in the composition arises from the
loss of sodium during the solid-state reaction. Fig. 2 shows the
XRD pattern of the as-quenched LNCO sample. All the XRD peaks can
be well indexed using a hexagonal cell with the P6$_{3}$/$mmc$
space group. The lattice parameters refined by a least squares fit
are $a$=2.824 \AA\ and $c$=20.293 \AA. It can be seen that the $c$
axis is taken as the expected value of
$2(c_{\text{LCO}}/3+c_{\text{NCO}}/2)$, where $c_{\text{LCO}}$ and
$c_{\text{NCO}}$ are the $c$ axis of the LCO O3
phase\cite{Levasseur} and that of the NCO P2
phase\cite{Fouassier}, respectively. One may also note that the
$a$ axis lies in between those of LCO\cite{Levasseur} (2.816 \AA)
and NCO\cite{Fouassier,Bayrakci} (2.830 \AA-2.843 \AA). This
implies that the oxygen sheets closed to the lithium layers are
under some compression, while those neighboring the sodium layers
bears some tension. Such asymmetric interactions on the conducting
CoO$_2$ layers might bring about some additional effects on
physical properties.

\begin{figure}[tbp]
\includegraphics[width=7cm]{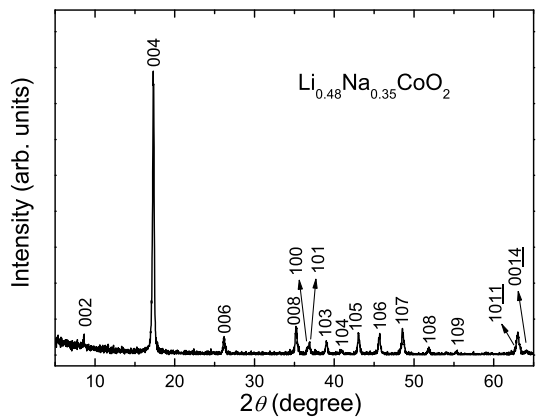}
\caption{X-ray powder diffraction pattern for the as-quenched
sample of Li$_{0.48}$Na$_{0.35}$CoO$_{2}$.}
\end{figure}

The measurements of electrical resistivity, thermal conductivity
and thermopower were performed on rectangular bars cut from the
annealed LNCO pellets. The resistivity was obtained using a
standard four-probe technique with a dc current of 0.5 mA. The
thermopower and the thermal conduction were measured by a
steady-state method with the temperature gradient of $\sim$2 K/cm.
All the electrodes were carefully made using well-conductive
silver paste. Fig. 3 shows the temperature dependence of
resistivity and thermal conductivity for the as-annealed LNCO
sample. The resistivity basically exhibits the semiconductor
behavior ($i$.$e$., $d$$\rho$/$dT$$<$0) in the whole temperature
range. The room temperature resistivity is $\sim$ 20 m$\Omega$ cm.
This value is about 6 times larger than that of
NaCo$_{2}$O$_{4}$\cite{Takahata} or
Na$_{0.75}$CoO$_{2}$\cite{Motohashi}, two orders of magnitude
smaller than that of Li$_{0.96}$CoO$_{2}$\cite{Levasseur}, and
comparable to that of Ca$_3$Co$_4$O$_9$\cite{Masset,Li}. It is
noted that the $\rho(T)$ curve shows an anomaly at $\sim$ 150 K
with a thermal hysteresis. We speculate that this phenomenon could
be related to the ordering of Na$^{+}$ ions (ordering of Li$^{+}$
ions is ignored because Li$^{+}$ almost fully occupies its lattice
site). The disorder of sodium ions at high temperatures induces
stronger Anderson localization effect, which might lead to in the
decrease in resistivity when cooling down to $\sim$ 150 K. It is
noted that ordering of sodium ions has been identified by electron
diffractions in NCO system\cite{Zandbergen}.

\begin{figure}[tbp]
\includegraphics[width=8cm]{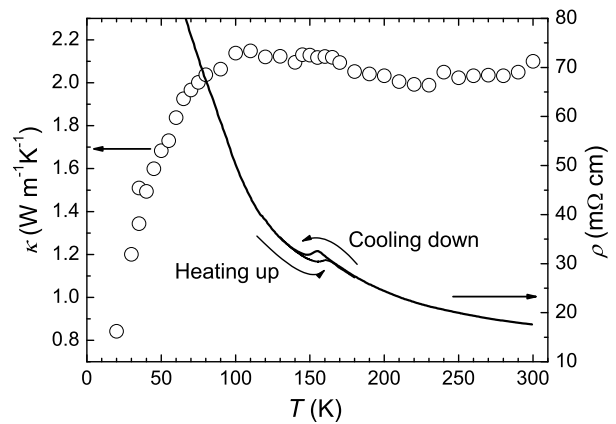}
\caption{Temperature dependence of electrical resistivity and
thermal conductivity for the as-annealed LNCO polycrystalline
sample.}
\end{figure}

Like NCO system, the thermal conductivity of the LNCO sample is
very low. The value of $\kappa$ at 300 K is $\sim$ 2 W m$^{-1}$
K$^{-1}$, close to that of the NCO polycrystalline
sample\cite{Takahata}. However, in contrast to NCO system which
shows that $\kappa$ increases obviously with increasing
temperature up to room temperature, $\kappa(T)$ of the LNCO sample
is weakly temperature dependent from 100 K to 300 K. Detailed
analysis of the $\kappa(T)$ behavior is beyond the scope of this
Letter. Nevertheless, we believe that the unusually low thermal
conductivity is also due to a phonon glass effect, which has been
proposed by Takahata et al.\cite{Takahata} for explaining the
thermal conductivity in NCO system.

Fig. 4 shows the thermopower as a function of temperature for the
identical sample. The positive sign of $S$ in the whole
temperature range establishes the hole-like character of the
prevailing charge carriers. Although the sample shows
semiconducting behavior in the resistivity measurement, the $S(T)$
curve exhibits the similar temperature dependence as the in-plane
thermopower of the metallic NCO crystals\cite{Terasaki}.
Strikingly, the thermopower at 300 K is as high as 182 $\mu$V/K,
which is $\sim 60\ \mu$V/K larger than the $S$ values of
Na$_{0.75}$CoO$_{2}$\cite{Motohashi} and
LiCoO$_{2}$\cite{Levasseur}. Therefore, the LNCO sample shows an
enhanced thermopower.

\begin{figure}[tbp]
\includegraphics[width=8cm]{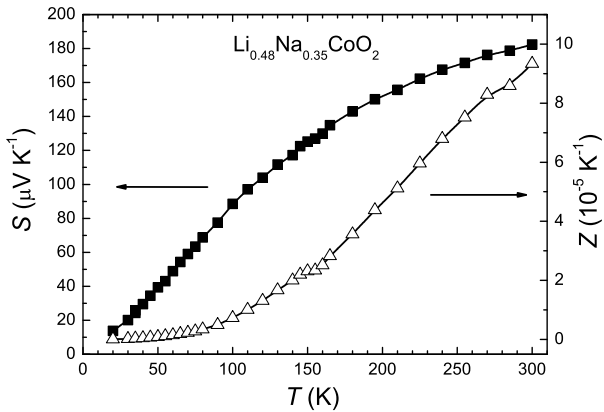}
\caption{Temperature dependence of thermopower ($S$) and figure of
merit ($Z$) for the as-annealed LNCO polycrystalline sample.}
\end{figure}

The origin of large thermopower in NCO has been investigated both
theoretically and experimentally for a few years. Koshibae et
al.\cite{Koshibae} considered that the degeneracy of carriers,
together with strong correlation of 3$d$ electrons, brought about
the large thermopower. Based on the LDA band calculation, on the
other hand, it was also able to give large thermopower since the
density of state $N(E)$ has a large slope at Fermi
level.\cite{Singh} Wang et al.\cite{Wang} studied the magnetic
field dependence of thermopower in NCO, and they concluded that
the spin entropy was the likely origin of the enhanced
thermopower. As for the LNCO system, in addition to the spin
entropy as one of the sources of the large thermopower, the
special intergrowth structure may also play a role in the
thermopower enhancement.

Using the above data of $S(T)$, $\rho(T)$ and $\kappa(T)$, the
figure of merit $Z$ ($=S^{2}/\rho\kappa$) can be obtained, as also
shown in Fig. 4. The $Z$ value at room temperature reaches nearly
10$^{-4}$ K$^{-1}$, which is better than or comparable to those of
other layered thermoelectric
cobaltites\cite{Takahata,Li,Funahashi}. Besides, the three factors
of $S(T)$, $\rho(T)$ and $\kappa(T)$ cooperatively raise the $Z$
value with further increasing temperature. Therefore, LNCO system
should be another promising candidate for high-temperature
thermoelectric applications.

This work was supported by the Specialized Research Fund for the
Doctoral Program of Higher Education (Grant No. 20040330563) and
the National Science Foundation of China (Grant No. 10225417).

\end{document}